\documentclass[10pt,a4paper]{article}
\usepackage[a4paper, total={6in, 8in}]{geometry} 

\usepackage[latin1]{inputenc}
\usepackage{amsfonts}
\usepackage{amssymb}
\usepackage{amsmath}
\usepackage{graphicx,psfrag,epsf}
\usepackage{booktabs}%
\usepackage{float}
\usepackage{subcaption} 

\usepackage{algorithm}%
\usepackage{algorithmicx}%
\usepackage{algpseudocode}%

\usepackage[round]{natbib}
\usepackage{hyperref}

\algrenewcommand\algorithmicrequire{\textbf{Input:}}
\algrenewcommand\algorithmicensure{\textbf{Output:}}

\newtheorem{theorem}{Theorem}
 
\newtheorem{proposition}[theorem]{Proposition} 

\newtheorem{corollary}[theorem]{Corollary}

\author{
Kes Ward\\
{k.ward4@lancaster.ac.uk}\\
Department of Mathematics and Statistics,\\ Lancaster University, UK\\\vspace{2px}\\
Gaetano Romano\\
{g.romano@lancaster.ac.uk}\\
Department of Mathematics and Statistics,\\ Lancaster University, UK\\\vspace{2px}\\
Idris Eckley\\
{i.eckley@lancaster.ac.uk}\\
Department of Mathematics and Statistics,\\ Lancaster University, UK\\\vspace{2px}\\
Paul Fearnhead\\
{p.fearnhead@lancaster.ac.uk}\\
Department of Mathematics and Statistics,\\ Lancaster University, UK\\\vspace{2px}\\
}
\title{A Constant-per-Iteration Likelihood Ratio Test for Online Changepoint Detection for Exponential Family Models}

\begin{document}
	
	\maketitle
	
	\begin{abstract}
		Online changepoint detection algorithms that are based on likelihood-ratio tests have been shown to have excellent statistical properties. However, a simple online implementation is computationally infeasible as, at time $T$, it involves considering $O(T)$ possible locations for the change. Recently, the FOCuS algorithm has been introduced for detecting changes in mean in Gaussian data that decreases the per-iteration cost to $O(\log T)$. This is possible by using pruning ideas, which reduce the set of changepoint locations that need to be considered at time $T$ to approximately $\log T$. We show that if one wishes to perform the likelihood ratio test for a different one-parameter exponential family model, then exactly the same pruning rule can be used, and again one need only consider approximately $\log T$ locations at iteration $T$. Furthermore, we show how we can adaptively perform the maximisation step of the algorithm so that we need only maximise the test statistic over a small subset of these possible locations. Empirical results show that the resulting online algorithm, which can detect changes under a wide range of models, has a constant-per-iteration cost on average.    
	\end{abstract}

	\section{Introduction}
	
	Detecting changes in data streams is an important statistical and machine learning challenge that arises in applications as diverse as climate records \cite[]{beaulieu2018distinguishing}, financial time-series \cite[]{andreou2002detecting}, monitoring performance of virtual machines \cite[]{barrett2017virtual} and detecting concept drift of inputs to classifiers \cite[]{sakamoto2015concept}. In many contemporary applications there is a need to detect changes online. In such settings we sequentially monitor a data stream over time, seeking to flag that a change has occurred as soon as possible. Often online change algorithms need to run under limited computational resource. For example, \cite{Ward2022} detect gamma ray bursts using the local computing resource onboard small cube satellites, and \cite{EdgeComputing} work with sensor networks where computations need to be performed locally by the sensors. Alternatively algorithms may need to be run for ultra high-frequency data \cite[]{iwata2018accelerating}, or need to be run concurrently across a large number of separate data streams. These settings share a common theme of tight constraints on the computational complexity of viable algorithms.
	
	
	There have been a number of procedures that have been suggested for online detection of changes, each involving different trade-offs between statistical efficiency and computational cost. For example, \cite{yu2020note} proposed a likelihood-ratio test with excellent statistical properties, but the natural implementation of this method has a computational cost per iteration that increases linearly with time. However, for online applications we need the computational cost to be constant. There exist algorithms with a constant computational cost per iteration, but they need one to only test for changes that are a pre-specified time in the past \cite[e.g.][]{eichinger2018mosum,Adams,Ross,chen2010modified}, or specify the distribution of the data after a change \cite[e.g.][]{Page,Lucas1985-gm}. If the choices made in implementing these algorithms are inappropriate for the actual change one wishes to detect, this can lead to a substantial loss of power. 
	
	
	Recently \cite{Romano2021} proposed a new algorithm called Functional Online Cumulative Sum (FOCuS). This algorithm is able to perform the likelihood-ratio test with a computational cost that only increases logarithmically with time. FOCuS was developed for detecting a change in mean in Gaussian data and has been extended to Poisson \cite[]{Ward2022} and Binomial \cite[]{Romano2022} data. FOCuS has two components: one that does pruning of past changepoint times that need not be considered in the future, and a maximisation step that considers all past changepoint times that have not been pruned. Interestingly, the pruning step for Poisson and Binomial data is identical to that for Gaussian data, and it is only the maximisation step that changes. 
	
	In this paper we show that this correspondence extends to other one-parameter exponential family models. Furthermore, we show how to substantially speed up FOCuS. In previous implementations the pruning step has a fixed average cost per iteration, and the computational bottleneck is the maximisation step that, at time $T$, needs to consider on average $O(\log T)$ possible changepoint locations. We show how previous calculations can be stored so that the maximisation step can consider fewer past changepoint locations. Empirically this leads to a maximisation step whose per iteration computational cost is $O(1)$. To our knowledge this is the first algorithm that exactly performs the likelihood-ratio test for detecting a change with an average constant-per-iteration cost.
	
	\section{Background}
	
	\subsection{Problem Statement}
	
	Assume we observe a univariate time series signal $x_1, x_2, ...$, and wish to analyse the data online and detect any change in the distribution of the data as quickly as possible. We will let $T$ denote the current time point. 
	
	A natural approach to this problem is to model the data as being independent realisations from some parametric family with density $f(x\mid\theta)$. Let $\theta_0$ be the parameter of the density before any change. If there is a change, denote the time of the change as $\tau$ and the parameter after the change as $\theta_1$. We can then test for a change using the likelihood-ratio test statistic. 
	
	There are two scenarios for such a test. First, we can assume the pre-change distribution, and hence $\theta_0$ is known \cite[]{eichinger2018mosum}. This simplifying assumption is commonly made when we have substantial training data from the pre-change distribution with which to estimate $\theta_0$. Alternatively, we can let $\theta_0$ be unknown. We will initially focus on the pre-change distribution known case, and explain how to extend ideas to the pre-change distribution unknown case in Section \ref{sec:theta0unknown}.
	
	The log-likelihood for the data $x_{1:T}=(x_1,\ldots,x_T)$, which depends on the pre-change parameter, $\theta_0$, the post-change parameter, $\theta_1$, and the location of a change, $\tau$, is
	\[
	\ell(x_{1:T} \vert \theta_0, \theta_1, \tau ) := \sum_{t=1}^{\tau} \log f(x_t \vert \theta_0) + \sum_{t=\tau+1}^{T} \log f(x_t \vert \theta_1).
	\]
	The log-likelihood ratio test statistic for a change prior to $T$ is thus
	$$LR_T := 2\left\{\max_{\theta_1, \tau} \ell(x_{1:T} \vert\theta_0, \theta_1, \tau) - \ell(x_{1:T} \vert \theta_0, \cdot, T) \right\}. $$
	Naively calculating the log-likelihood ratio statistic involves maximising over a set of $T$ terms at time $T$. This makes it computationally prohibitive to calculate in an online setting when $T$ is large. There are two simple pre-existing approaches to overcome this, and make the computational cost per iteration constant. First, MOSUM approaches \cite[e.g.][]{chu1995mosum,eichinger2018mosum} fix a number, $K$ say, of changepoint times to be tested, with these being of the form $\tau=T-h_i$ for a suitable choice of $h_1,\ldots,h_K$. Alternatively one can use Page's recursion \cite[]{Page,page1955test} that calculates the likelihood-ratio test statistic for a pre-specified post-change parameter. Again we can use a grid of $K$ possible post-change parameters. Both these approaches lose statistical power if the choice of either changepoint location (i.e.\ the $h_i$ values for MOSUM) or the post-change parameter are inappropriate for the actual change in the data we are analysing.

	\subsection{FOCuS for Gaussian data} \label{sec:FOCuS}
	
	As an alternative to MOSUM or Page's recursion, \cite{Romano2021} introduce the FOCuS algorithm that can efficiently calculate the log-likelihood ratio statistic for univariate Gaussian data where $\theta$ denotes the data mean. 
	
	In this setting. it is simple to see that 
	\begin{eqnarray*}
		\ell(x_{1:T} \left\vert \theta_0, \theta_1, \tau \right. ) - \ell(x_{1:T} \vert \theta_0, \cdot, T) = \\  \sum_{t=\tau+1}^T \left\{\log f(x_t\vert\theta_1)-\log f(x_t\vert \theta_0)\right\}.
	\end{eqnarray*}
	We can then introduce a function 
	\[
	Q_T(\theta_1)=\max_\tau \left\{
	\sum_{t=\tau+1}^T \Big(\log f(x_t\vert\theta_1)-\log f(x_t\vert \theta_0)\Big) 
	\right\},
	\]
	which is the log-likelihood ratio statistic if the post-change parameter, $\theta_1$, is known. Obviously, $LR_T=\max_{\theta_1} 2Q_T(\theta_1)$.
	
	For Gaussian data with known mean, $\theta_0$, and variance, $\sigma^2$, we can standardise the data so that the pre-change mean is 0 and the variance is 1. In this case, each term in the sum of the log-likelihood ratio statistic simplifies to $\theta_1(x_t-\theta_1/2)$, and
	\begin{equation} \label{eq:1}
		Q_T(\theta_1)=\max_\tau \left\{
		\sum_{t=\tau+1}^T \theta_1(x_t-\theta_1/2)
		\right\}.
	\end{equation}
	This is the point-wise maximum of $T-1$ quadratics. We can thus store $Q_t(\theta_1)$ by storing the coefficients of the quadratics.
	
	The idea of FOCuS is to recursively calculate $Q_T(\theta_1)$. Whilst we have written  $Q_T(\theta_1)$ as the maximum of $T-1$ quadratics in $\theta_1$, each corresponding to a different location of the putative change, in practice there are only $\approx \log T$ quadratics that contribute to $Q_T$ \cite[]{Romano2021}. This means that, if we can identify this set of quadratics,  we can maximise $Q_T$, and hence calculate the test statistic, in $O(\log T)$ operations. Furthermore \cite{Romano2021} show that we can recursively calculate $Q_T$, and the minimal set of quadratics we need, with a cost that is $O(1)$ per iteration on average. 
	
	The FOCuS recursion is easiest described for the case where we want a positive change, i.e. $\theta_1>\theta_0$. An identical recursion can then be applied for $\theta_1<\theta_0$ and the results combined to get $Q_T$.
	This approach to calculating $Q_T$ uses the recursion of \cite{Page}, 
	\[
	Q_T(\theta_1) = \max\left\{Q_{T-1}(\theta_1),0 \right\} + \theta_1(x_T-\theta_1/2).
	\]
	To explain how to efficiently solve this recursion, it is helpful to introduce some notation. For $\tau_i < \tau_j$ define 
	\begin{equation} \label{eq:C_s^t}
		\mathcal{C}_{\tau_i}^{(\tau_j)}(\theta_1) = \sum_{t=\tau_i+1}^{\tau_j} \theta_1(x_t-\theta_1/2).
	\end{equation}
	At time $T-1$ let the quadratics that contribute to $Q_{T-1}$, for $\theta_1>\theta_0$, correspond to changes at times $\tau\in \mathcal{I}_{T-1}$. Then
	\[
	Q_{T-1}(\theta_1) = \max_{\tau\in\mathcal{I}_{T-1}} \left\{ 
	\mathcal{C}_{\tau}^{(T-1)}(\theta_1)
	\right\}.
	\]
	Substituting into Page's recursion we obtain
	\[
	Q_T(\theta_1) = \max\left\{ \max_{\tau\in\mathcal{I}_{T-1}} \left\{ 
	\mathcal{C}_{\tau}^{(T)}(\theta_1)
	\right\}, \mathcal{C}_{T-1}^T(\theta_1)
	\right\},
	\]
	from which we have that $\mathcal{I}_T\subseteq \mathcal{I}_{T-1}\cup \{T-1\}.$ 
	
	The key step now is deciding which changepoint locations in $\mathcal{I}_{T-1}\cup \{T-1\}$ no longer contribute to $Q_T$. To be consistent with ideas we present in Section \ref{sec:FOCuS_expo_fam} we will present the FOCuS algorithm in a slightly different way to \cite{Romano2021}. Assume that $\mathcal{I}_{T-1}=\{\tau_1,\ldots,\tau_n\}$, with the candidate locations ordered so that $\tau_1<\tau_2<\ldots<\tau_n$.  We can now define the difference between successive quadratics as
	\begin{eqnarray*}
		\mathcal{C}_{\tau_i}^{(T)}(\theta_1)-\mathcal{C}_{\tau_{i+1}}^{(T)}(\theta_1)&=&
		\mathcal{C}_{\tau_i}^{(T-1)}(\theta_1)-\mathcal{C}_{\tau_{i+1}}^{(T-1)}(\theta_1) \\
		&=&\mathcal{C}_{\tau_i}^{(\tau_{i+1})}(\theta_1).
	\end{eqnarray*}
	These differences do not change from time $T-1$ to time $T$.
	
	For the difference between quadratics associated with changes at $\tau_i$ and $\tau_{i+1}$, let $l_i\geq 0$ denote the largest value of $\theta_1$ such $\mathcal{C}_{\tau_i}^{(\tau_{i+1})}(\theta_1)\geq0$. By definition $\mathcal{C}_{\tau_i}^{(\tau_{i+1})}(\theta_0)=0$. Hence it is readily shown that
	\[
	\mathcal{C}_{\tau_i}^{(T)}(\theta_1) \geq \mathcal{C}_{\tau_{i+1}}^{(T)}(\theta_1),
	\]
	on $\theta\in[\theta_0,l_i]$. For $\theta_1\geq l_i$ compare $\mathcal{C}_{\tau_{i+1}}^{(T)}(\theta_1)$ with $\mathcal{C}_{T-1}^{(T)}(\theta_1)$. If
	$\mathcal{C}_{\tau_{i+1}}^{(T)}(\theta_1) \leq \mathcal{C}_{T-1}^{(T)}(\theta_1)$ then
	\begin{eqnarray*}
		\mathcal{C}_{\tau_{i+1}}^{(T)}(\theta_1) - \mathcal{C}_{T-1}^{(T)}(\theta_1) &\leq& 0 \\
		\Leftrightarrow \mathcal{C}_{\tau_{i+1}}^{(T-1)}(\theta_1) &\leq& 0.
	\end{eqnarray*}
	A sufficient condition for $\mathcal{C}_{\tau_{i+1}}^{(T-1)}(\theta_1) \leq 0$ for all $\theta_1>l_i$ is for the largest root of 
	$\mathcal{C}_{\tau_{i+1}}^{(T-1)}(\theta_1)$ to be smaller than $l_i$. In this case we have that $\mathcal{C}_{\tau_{i+1}}^{(T)}(\theta_1)$ does not contribute to $Q_T(\cdot)$ and thus can be pruned.
	
	
	This suggests Algorithm \ref{alg:FOCuS}. Note that this algorithm is presented differently from that in \cite{Romano2021}, as the way the quadratics are stored is different. Specifically, here we store the difference in the quadratics, rather than use summary statistics. The input is just the difference of the quadratics that contribute to $Q_{T-1}$.  The main loop of the algorithm just checks whether the root of $\mathcal{C}_{\tau_j}^{(T-1)}$ is smaller than that of $\mathcal{C}_{\tau_{j-1}}^{(\tau_j)}$, which is our condition for pruning the quadratic associated with $\tau_j$.  If not, we stop any further pruning and return the set of quadratic differences plus the quadratic $\mathcal{C}_{T-1}^{(T)}$. If it is, then the quadratic associated with $\tau_j$ is removed and  the quadratic difference associated with $\tau_{j-1}$ is updated -- by adding on the quadratic difference associated with $\tau_j$. We then loop to consider removing the next quadratic (if there is one). 
	
	\begin{algorithm}[ht]
		\caption{FOCuS update at time $T$ for $\theta_1>\theta_0$ and $\theta_0=0$. Algorithm based on storing quadratic differences.}
		\label{alg:FOCuS}
		\begin{algorithmic}[1]
			\Require     A set of $n$ quadratic differences, $\mathcal{C}_{\tau_i}^{(\tau_{i+1})}(\theta_1)$, for $i=1,\ldots,n$, with $\tau_i<\tau_{i+1}$ and $\tau_{n+1}=T-1$ such that 
			\[
			Q_{T-1}(\theta_1)=\max_{i}\{\mathcal{C}_{\tau_i}^{(\tau_{i+1})}\}.
			\]
			The set of largest roots, $l_i$, such that $\mathcal{C}_{\tau_i}^{(\tau_{i+1})}(l_i)=0$, for $i=1,\ldots,n$.
			
			\vspace*{1.0em}
			
			\State $j\gets n$
			\State $l_0\gets \theta_0$
			\While{$j>0$}
			\If{ $l_j\leq l_{j-1}$}
			\State $C_{\tau_{j-1}}^{(T-1)}(\theta_1)\gets C_{\tau_{j-1}}^{\tau_j}(\theta_1)+C_{\tau_{j}}^{(T-1)}(\theta_1)$
			\State Recalculate $l_{j-1}$, largest root of $C_{\tau_{j-1}}^{(T-1)}(\theta_1)=0$
			\State $\tau_j\gets T-1$
			\State $j\gets j-1$
			\EndIf
			\State Break
			\EndWhile
			\vspace*{0.5em}
			\State $\mathcal{C}_{T-1}^{(T)}(\theta_1)\gets\theta_1(x_T-\theta_1/2)$
			\State $\tau_{j+1}\gets T-1$ and $\tau_{j+2}\gets T$
			\State $l_{j+1}\gets 2x_T$
			\State $n\gets j+1$
			\Ensure The set of $n$ quadratic differences, $\mathcal{C}_{\tau_i}^{(\tau_i+1)}(\theta_1)$ and roots $l_i$ for $i=1,\ldots,n$.
		\end{algorithmic}
		
		\vspace*{1.0em}
		
	\end{algorithm}
	
	\begin{figure}
		\centering
		\includegraphics[width=\linewidth]{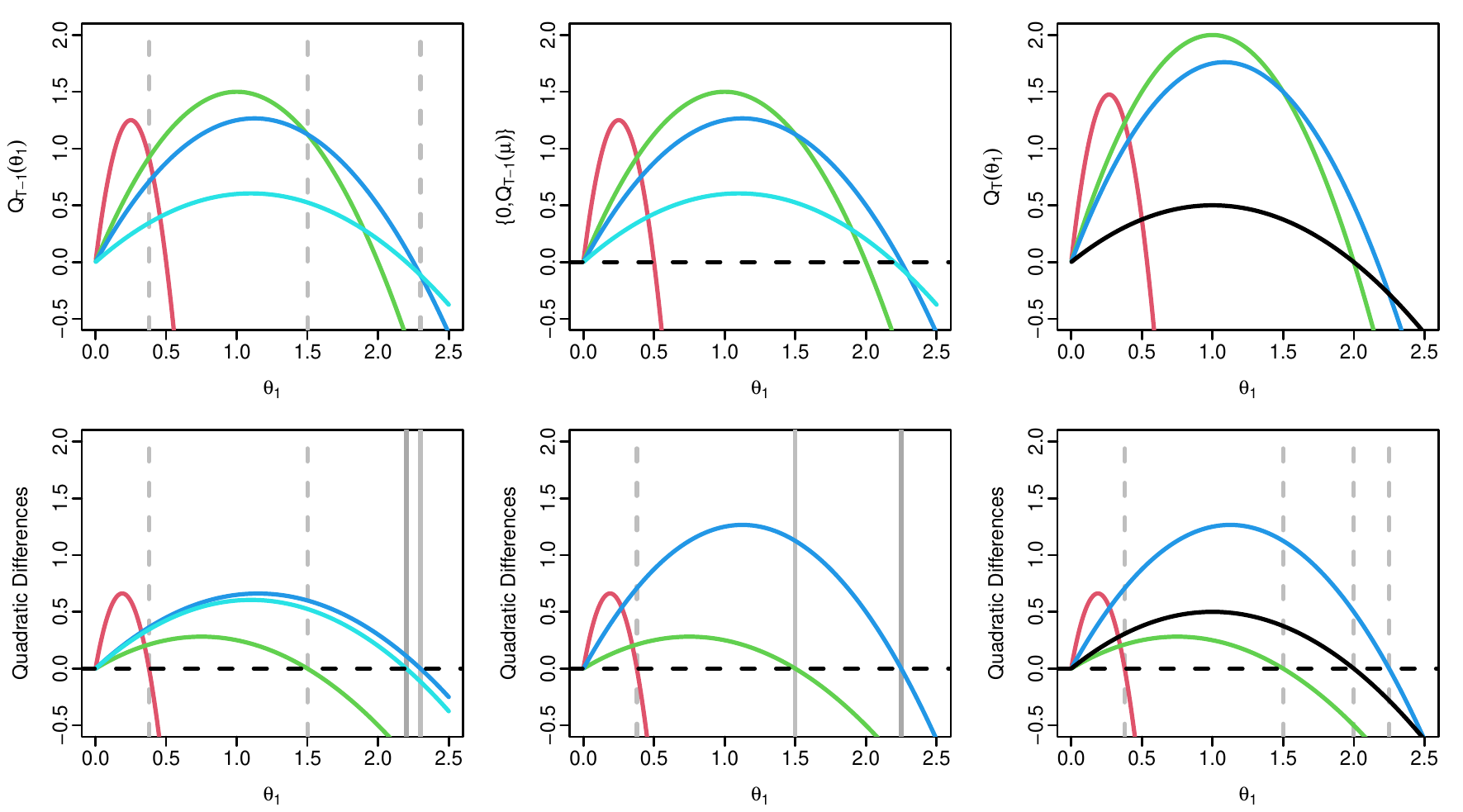}
		\caption{Example of one iteration of FOCuS. The top row plots the quadratics $C_{\tau_{1}}^{(T-1)}$ (red), $C_{\tau_{2}}^{(T-1)}$ (green), $C_{\tau_{3}}^{(T-1)}$ (blue), $C_{\tau_{4}}^{(T-1)}$ (cyan) that contribute to $Q_T(\theta_1)$ directly, together with the intervals where each is optimal (demarked by grey vertical lines). To prune, we first add the zero line  (dotted black), then prune $C_{\tau_{4}}^{(T-1)}$, as it is no longer optimal for any $\theta_1$. We then add $
			\theta_1(x_T-\theta_1/2)$ to all quadratics. The bottom-left plot shows the storage of quadratic differences $C_{\tau_{1}}^{(\tau_2)}$ (red), $C_{\tau_{2}}^{(\tau_3)}$ (green), $C_{\tau_{3}}^{(\tau_4)}$ (blue), $C_{\tau_{4}}^{(T-1)}$ (cyan) in Algorithm \ref{alg:FOCuS}. The roots of these quadratic differences are shown by grey vertical lines. The roots of the first three quadratic difference demark the intervals where the quadratics are optimal. The root of $C_{\tau_{4}}^{(T-1)}$ shows the region where that curve is above the zero-line. The algorithm considers pruning $\tau_4$ based on whether the root of  $C_{\tau_{4}}^{(T-1)}$ is smaller than the root of $C_{\tau_{3}}^{(\tau_4)}$. The pruning of $\tau_4$ 
			combines cyan with blue into the quadratic difference $C_{\tau_{3}}^{(T-1)}$ (bottom-middle, blue line). We then add $C_{T-1}^{(T)}$ (black) as its own quadratic difference (bottom-right). We require no iteration over the full quadratic list, as $C_{\tau_{1}}^{(\tau_2)}$ and $C_{\tau_{2}}^{(\tau_3)}$ remain untouched. 
		}
		\label{fig:FOCUSexample}
	\end{figure}
	
	A pictorial description of the algorithm is shown in Figure \ref{fig:FOCUSexample}. It is simple to see that this algorithm has an average cost per iteration that is $O(1)$. This is because,  at each iteration, the number of steps of the while loop is one more than the number of quadratics that are pruned. As only one quadratic is added at each iteration, and a quadratic can only be removed once, the overall number of steps of the while loop by time $T$ will be less than $2T$.
	
	\section{FOCuS for Exponential Family Models}\label{sec:FOCuS_expo_fam}
	
	Different parametric families will have different likelihoods, and likelihood ratio statistics. However the idea behind FOCuS can still be applied in these cases provided we are considering a change in a univariate parameter, with different forms for the curves (described in Equation \ref{eq:C_s^t}) and hence different values for the roots of the curves. Whilst one would guess that the different values of the roots would lead to different pruning of curves when implementing Algorithm \ref{alg:FOCuS}, \cite{Ward2022} and \cite{Romano2022} noted that the pruning, i.e. the changepoints associated with the functions that contribute to $Q_T$, are the same for a Poisson model or a Binomial model as for the Gaussian model; it is only the shape of the functions that changes. Here we show that this is a general property for many one-parameter exponential family models.
	
	A one-parameter exponential family distribution can be written as
	
	$${f(x\mid \theta )=\exp \!{\bigl [}\,\alpha (\theta )\cdot \gamma(x)-\beta(\theta )+\delta(x)\,{\bigr ]}},$$
	
	for some one-parameter functions $\alpha(\theta), \beta(\theta), \gamma(x), \delta(x)$ which are dependent on the specific distribution. Examples of one-parameter exponential family distributions given in Table \ref{table:exponential_families} include Gaussian change in mean, Gaussian change in variance, Poisson, Gamma change in scale, and Binomial distributions, for which $\alpha(\theta)$ and $\beta(\theta)$ are increasing functions. $\gamma(x)$ is the sufficient statistic for the model, and is often the identity function. We do not need to consider $\delta(x)$ as it cancels out in all likelihood ratios.
	
	There are various simple transformations that can be done to shift data points from one assumed exponential family form to another before applying change detection methods, for example binning Exponentially distributed data into time bins to give rise to Poisson data, approximating Binomal($n, \theta$) data as $\text{Poisson}(n\theta)$ for large $n$ and small $\theta$, or utilising the fact that $x \sim N(0, 1)$ then $x^2 \sim \text{Gamma}(1/2, 1/2)$ to turn a Gaussian change in variance problem into a Gamma change in parameter problem (refer to Section \ref{sec:simulations} for an illustration of this). Nevertheless, the ability to work flexibly in all possible exponential family settings without requiring data pre-processing can be helpful.
	
	\begin{table*}[tb]
		\centering
		\begin{tabular}{llll}
			\hline
			Distribution                    & $\alpha(\theta)$                  & $\beta(\theta)$           & $\gamma(x)$\\ \hline
			Gaussian (change in mean)       & $\theta$                          & $\theta^2$                & $x$  \\
			Gaussian (change in variance)   & $-1/\theta^2$                      & $\log(\theta)$           & $x^2$ \\
			Poisson                         & $\log(\theta)$                    & $\theta$                  & $x$  \\
			Binomial                        & $\log(\theta)-\log(1\!-\!\theta)$ & $-n\log(1\!-\!\theta)$    & $x$  \\
			Gamma                           & $-1/\theta$                       & $k\log(\theta)$           & $x$  \\ \hline
		\end{tabular}%
		
		\caption{Examples of one-parameter exponential families and the corresponding forms of $\alpha(\theta)$, $\beta(\theta)$ and $\gamma(x)$. The Gaussian change in mean model is for a variance of 1, the Gaussian change in variance model is for a mean of 0; the Binomial model assumes the number of trials is $n$; and the Gamma model is for a change in scale parameter with shape parameter $k$.}
		\label{table:exponential_families}
	\end{table*}
	
	The ideas from Section \ref{sec:FOCuS} can be applied to detecting a change in the parameter of a one-parameter exponential family. The main change is to the form of the log-likelihood. For Algorithm \ref{alg:FOCuS} we need to store the differences $C_{\tau_i}^{(\tau_j)}(\theta_1)$ in the log-likelihood for different choices of the changepoint location. This becomes
	\begin{align*}
		\ell(x_{1:T} & \vert\theta_0, \theta_1, \tau_i) - \ell(x_{1:T} \vert\theta_0, \theta_1, \tau_j)  = \\
		= & \ [\alpha (\theta_1)-\alpha (\theta_0)]\sum_{t=\tau_i+1}^{\tau_j}\gamma(x_t) - [\beta(\theta_1)-\beta(\theta_0)](\tau_j-\tau_i). 
	\end{align*}
	These curves can summarised in terms of the coefficients of $\alpha(\theta_1)-\alpha(\theta_0)$ and $\beta(\theta_1)-\beta(\theta_0)$, that is $\sum_{t=\tau_i+1}^{\tau_j}\gamma(x_t)$ and $\tau_j-\tau_i$.
	
	The pruning of Algorithm \ref{alg:FOCuS} is based on comparing roots of curves. One challenge with implementing the algorithm for general exponential family models is that the roots are often not available analytically, unlike for the Gaussian model, and thus require numerical root finders.  However, pruning just depends on the ordering of the roots. The following proposition shows that we can often determine which of two curves has the larger root without having to calculate the value of the root. 
	
	Define 
	\[
	\bar{\gamma}_{\tau_i:\tau_j}=\frac{1}{\tau_j-\tau_i} \sum_{t=\tau_i+1}^{\tau_j} \gamma(x_t),
	\]
	to be the average value of $\gamma(x_t)$ for $t=\tau_i+1,\ldots,\tau_j$, and define $\theta_1^\tau(\neq \theta_0)$ to be the root of
	\[
	\ell(x_{1:T} \vert\theta_0, \theta_1^\tau, \tau) - \ell(x_{1:T} \vert\theta_0, \cdot, T) = 0.
	\]
	Then the following proposition shows that the ordering of the roots is determined by the ordering of $\bar{\gamma}$ values.
	
	\begin{proposition} \label{prop:1}
		Suppose that for our choice of $\theta_0$ the function
		
		\[\theta_1 : \rightarrow \frac{\beta(\theta_1)-\beta(\theta_0)}{\alpha(\theta_1)-\alpha(\theta_0)} \]
		is strictly increasing. Then the sign of $\bar{\gamma}_{\tau_i:\tau_j} - \bar{\gamma}_{\tau_j:T}$ is the same as the sign of $\theta_1^{\tau_i} - \theta_1^{\tau_j}$.
	\end{proposition}
	
	{\bf Proof:} See Supplementary Material.
	
	\noindent In other words, $\theta_1^{\tau_i} > \theta_1^{\tau_j}$ if and only if $\bar{\gamma}_{\tau_i:\tau_j} > \bar{\gamma}_{\tau_j:T}$. Thus we can change the condition in Algorithm \ref{alg:FOCuS} that compares the roots of two curves with a condition that compares their $\bar{\gamma}$ values. Or equivalently we can implement Algorithm \ref{alg:FOCuS} but with $l_i=\bar{\gamma}_{\tau_i:\tau_{i+1}}$ rather than the root of $\mathcal{C}_{\tau_i}^{\tau_{i+1}}=0$.
	
	An immediate consequence of this result is that one-parameter exponential family models that satisfy the condition of Proposition \ref{prop:1} and that have the same value for $\gamma(x)$ will prune exactly the same set of curves. 
	This leads to the following corollary based on a set of exponential family models with $\gamma(x)=x$, the same as the Gaussian change in mean model of the original FOCuS algorithm.

	\begin{corollary}
		The Gaussian (change in mean), Poisson, Binomial, and Gamma variations of the FOCuS algorithm have the same pruning.
	\end{corollary}
	
\begin{figure}[tb]
		\centering
		\includegraphics[width=\textwidth]{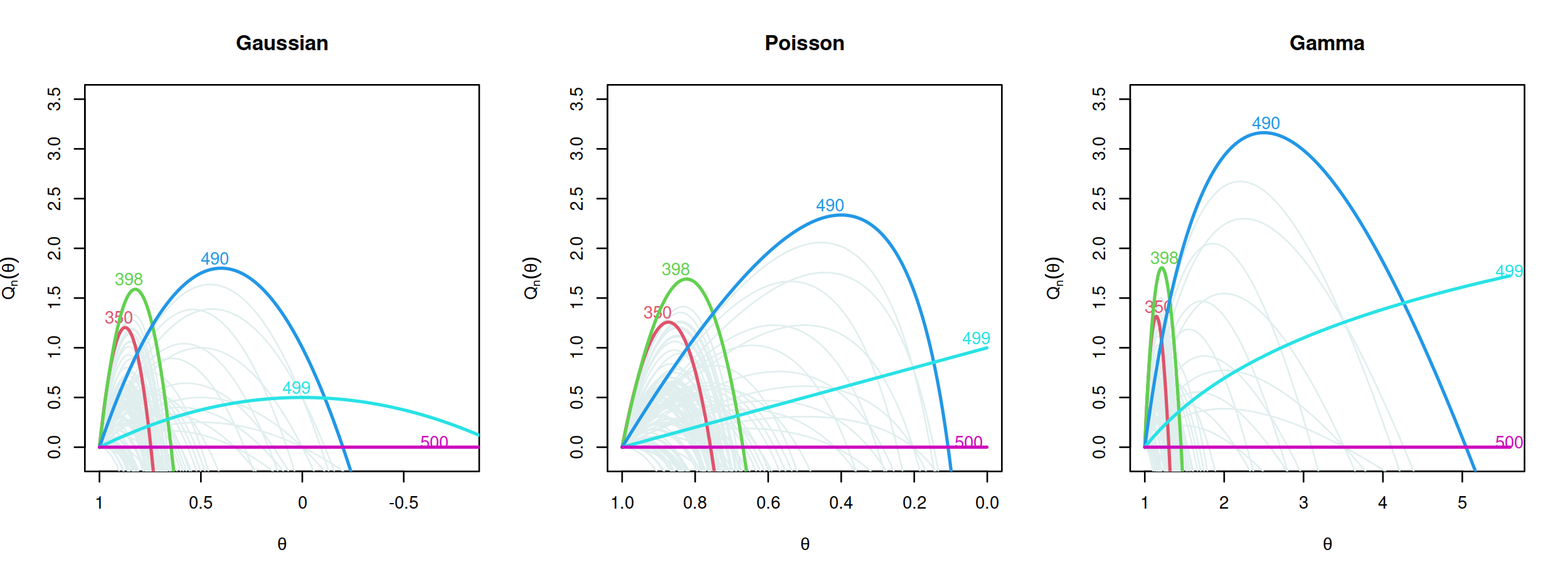}
		\caption{Comparison of three different cost functions computed from the same realizations $y_1, \dots, y_{500} \sim \text{Poi}(1)$. The leftmost, center, and rightmost figures show the cost function $Q_n(\theta)$ should we assume respectively a Gaussian, Poisson, or Gamma loss. The floating number refers to the timestep at which each curve was introduced. In gray, the curves that are no longer optimal and hence were pruned.}
		\label{fig:same_timesteps}
	\end{figure}

	A graphical example of this corollary is shown in Figure \ref{fig:same_timesteps}.
	
	More generally we have the following.
	\begin{corollary} \label{cor:2}
		If an exponential family model satisfies the condition of Proposition \ref{prop:1}, then the pruning under this model will be identical to the pruning of FOCuS for the Gaussian change in mean model analysing data $\gamma(x_t)$.
	\end{corollary}
	So, for example, the pruning for the Gaussian change in variance model will be the same as for the Gaussian change in mean model run on data $x_1^2,x_2^2, \dots$.
	
	One consequence of this corollary is that the strong guarantees on the number of curves that are kept at time $T$ for the original FOCuS algorithm \cite[]{Romano2021} applies to these equivalent exponential family models. The results on the expected number of curves kept by FOCuS makes minimal assumptions for the data, namely that the observations are exchangeable. These results imply the on average the number of curves kept at iteration $T$ is $O(\log T)$.
	
	

	
	
	


	\section{Unknown Pre-change Parameter}
	\label{sec:theta0unknown}
	
	We next turn to consider how to extend the methodology to the case where both pre-change and post-change parameters are unknown. When $\theta_0$ is unknown, the log likelihood-ratio statistic, $LR_T$, satisfies
	\begin{eqnarray*}
		\frac{LR_T}{2} &= & \max_{\theta_0,\theta_1, \tau} \left\{
		\sum_{t=1}^\tau \log f(x_t\vert\theta_0)+\sum_{t=\tau+1}^T \log f(x_t\vert\theta_1)
		\right\} \\
		& & - \max_{\theta_0}\sum_{t=1}^T \log f(x_t\vert\theta_0) .
	\end{eqnarray*}
	The challenge with calculating this is the first term. Define
	\[
	Q^*_T(\theta_0,\theta_1)=\max_{\tau}\left\{ \sum_{t=1}^\tau \log f(x_t\vert\theta_0)+\sum_{t=\tau+1}^T \log f(x_t\vert\theta_1)\right\}.
	\]
	If we can calculate this function of $\theta_0$ and $\theta_1$, it will be straightforward to calculate the likelihood-ratio statistic. If we fix $\theta_0$ and consider $Q^*_T$ as a function of only $\theta_1$ then this is just the function $Q_T(\theta_1)$ we considered in the known pre-change parameter.
	
	As before, we can write $Q^*_T(\theta_0,\theta_1)$ as the maximum of a set of curves, now of two variables $\theta_0$ and $\theta_1$, with each function relating to a specific value of $\tau$. As before if we can easily determine the curves for which values of $\tau$ contribute to the maximum, we can remove the other functions and greatly speed-up the calculation of $Q^*_T$.
	
	To do this, consider $Q^*_T(\theta_0,\theta_1)$ as a function of $\theta_1$ only, and write this as $Q_{T,\theta_0}(\theta_1)$. Algorithm \ref{alg:FOCuS} gives us the curves the contribute to this function for $\theta_1>\theta_0$. This set of curves is determined by the ordering of the roots of the curves, i.e. the $l_i$ for $i\geq 1$ in Algorithm \ref{alg:FOCuS}. If we now change $\theta_0$, the roots of the curves will change, but by Proposition \ref{prop:1} the orderings will not. The only difference will be with the definition of $l_0$. That is as we reduce $\theta_0$ we may have additional curves that contribute to the maximum, due to allowing a larger range of values for $\theta_1$, but as we increase $\theta_0$ we can only ever remove curves. I.e. we never swap the curves that need to be kept. Thus if we run Algorithm \ref{alg:FOCuS} for $\theta_0=-\infty$, then the set of curves we keep will be the set of curves that contribute to $Q^*_T(\theta_0,\theta_1)$ for $\theta_1>\theta_0$.
	
	
	
	
	
	
	
	In practice, this means that to implement the pruning of FOCuS with pre-change parameter unknown, we proceed as in Algorithm \ref{alg:FOCuS} but set $l_0 = -\infty$ when considering changes $\theta_1 > \theta_0$, and $l_0 = \infty$ when considering changes $\theta_1 < \theta_0$. The equivalence of Algorithm \ref{alg:FOCuS} across different exponential family models, that we demonstrated with  Corollary \ref{cor:2}, also immediately follows.
	
	\section{Adaptive Maxima Checking}\label{sec:apt_max_check}
	
	The main computational cost of the FOCuS algorithm comes from maximising the curves at each iteration. This is particularly the case for non-Gaussian models, as maximising a curve requires evaluating $\max_{\theta_0, \theta_1}\ell(x_{1:T}\vert\theta_0, \theta_1, \tau)$, which involves computing at least one logarithm (as in the cases of Poisson, Binomial, Gamma data). As the number of curves kept by time $T$ is of order $\log(T)$, calculating all maxima represents a (slowly) scaling cost. However we can reduce this cost by using information from previous iterations so that we need only maximise over fewer curves in order to detect whether $Q_T$ is above or below our threshold. This is possible by obtaining an upper bound on $Q_T$ that is easy to evaluate, as if this upper bound is less than our threshold we need not calculate $Q_T$.
	
\begin{figure}[!tb]
		\centering
		\includegraphics[width=.5\linewidth]{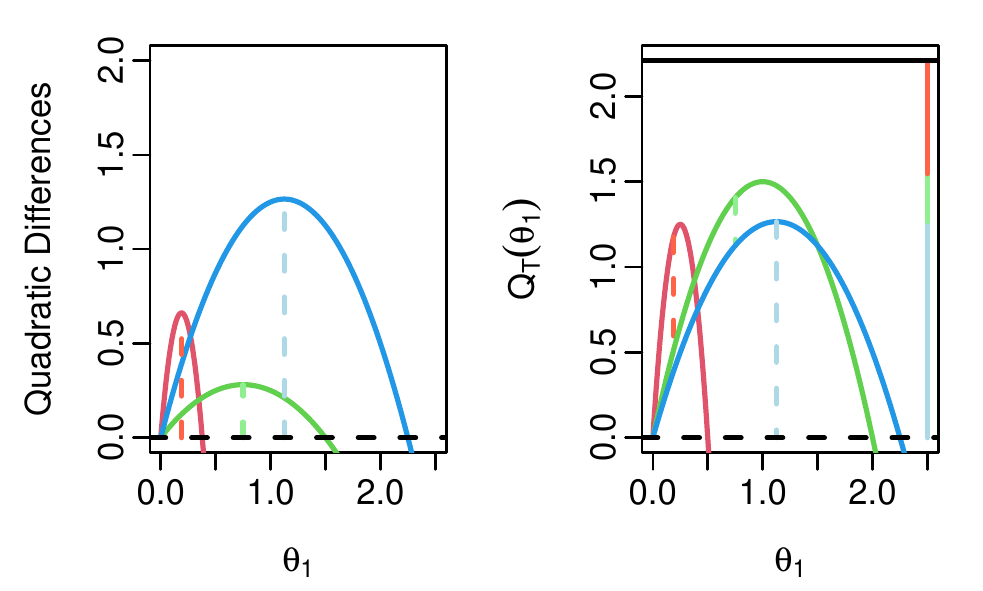}
		\caption{Example of the bound of Propositon \ref{prop:1} for the pre-change mean known case. Left-hand plot shows the differences between the three curves that contribute to $Q_T(\theta_1)$. The $m_{\tau_i:\tau_j}$ values correspond to the maximum of these curves (vertical lines). Right-hand plot shows $Q_{T}(\theta)$, the three curves that define it, and the maximum difference between the curves (vertical bars). The bound is the sum of the maximum differences (right-most stacked line). 
		}
		\label{fig:FOCUS_maxima}
	\end{figure}

	The following proposition gives such an upper bound on the maximum of all, or a subset, of curves. First for $\tau_i < \tau_j$, we define the likelihood ratio statistic for a change at $\tau_i$ with the signal ending at $\tau_j$. Define this likelihood ratio statistic as
	\begin{equation*}
		m_{\tau_i, \tau_j} = \max_{\substack{\theta_0\in H_0, \\ \theta_1}} \ell(x_{1:\tau_j} \vert\theta_0, \theta_1, \tau_i) - \max_{\theta_0\in H_0}\ell(x_{1:\tau_j} \vert\theta_0, \cdot, \tau_j),
	\end{equation*}
	where $H_0$ denotes the set of possible values of $\theta_0$. $H_0$ will contain a single value in the pre-change parameter known case, or be $\mathbb{R}$ for the pre-change parameter unknown case.
	\begin{proposition} \label{prop:2}
		For any $\tau_1 < \tau_2 < ... < \tau_n < T$, we have 
		\[
		\max_{i=1,...,n} m_{\tau_i, T} \leq \sum_{i=1}^{n-1} m_{\tau_i, \tau_{i+1}} + m_{\tau_n, T}.
		\]
	\end{proposition}	
	{\bf Proof:} See Supplementary Material. A pictorial explanation of the result is also shown in Figure \ref{fig:FOCUS_maxima}
	
	We can use this result as follows. The sum $M_{\tau_k} := \sum_{i=1}^{k-1} m_{\tau_i, \tau_{i+1}}$ can be stored as part of the likelihood curve for $\tau_k$, and the maxima checking step can proceed as in Algorithm \ref{alg:FOCuS_maxima}. The idea is that we can bound $Q_T$ above by $m_{\tau_k,T}+M_{\tau_k}$. So, starting with the curve with largest $\tau_k$ value we check if $m_{\tau_k,T}+M_{\tau_k}$ is below the threshold. If it is, we know $Q_T$ is below the threshold and we can output that no change is detected without considering any further curves. If not, we see if $m_{\tau_k,T}$, the likelihood-ratio test statistic for a change at $\tau_k$ is above the threshold. If it is we output that a change has been detected. If not then we proceed to curve with the next largest $\tau_k$ value and repeat.
	
	Empirical results suggest that for $\tau_1 ... \tau_n \in \mathcal{I}_T$ when searching only for an up-change (or analogously only for a down-change), the upper bound in Proposition \ref{prop:2} is quite tight under the underlying data scenario of no change because most of the $m_{\tau_i, \tau_{i+1}}$ are very small. Furthermore, as we show in Section \ref{sec:simulations}, at the majority of time-steps only one curve needs to be checked before we know that $Q_T$ is less than our threshold.
	
	\begin{algorithm}[tb]
		\caption{FOCuS maxima check at time $T$ for $\theta_1\geq\theta_0$.}
		\label{alg:FOCuS_maxima}
		\begin{algorithmic}[1]
			\Require{A set of $n$ likelihood curves and associated $(\tau_k, M_{\tau_k})$ values.}
			\vspace*{1em}
			
			\State Set $k=n$
			\While{$k>0$}
			\State Calculate $m_{\tau_k, T}$
			\If{ $m_{\tau_k, T} + M_{\tau_k} < \textit{Threshold}$}
			\State \textbf{break}
			\Else
			\State \textbf{Output:} Return change on $[\tau_k, T]$
			\EndIf
			\State $k \gets k-1$
			\EndWhile
			\Ensure Return no change.
		\end{algorithmic}
		\vspace*{1em}
	\end{algorithm}

	\section{Numerical Examples}
	
	\label{sec:simulations}
	We run some examples to empirically evaluate the computational complexity of the FOCuS procedure, comparing the various implementations presented in this paper with those already present in the literature. 
	
\begin{figure}[tb]
		\centering
		\includegraphics[width=0.5\linewidth]{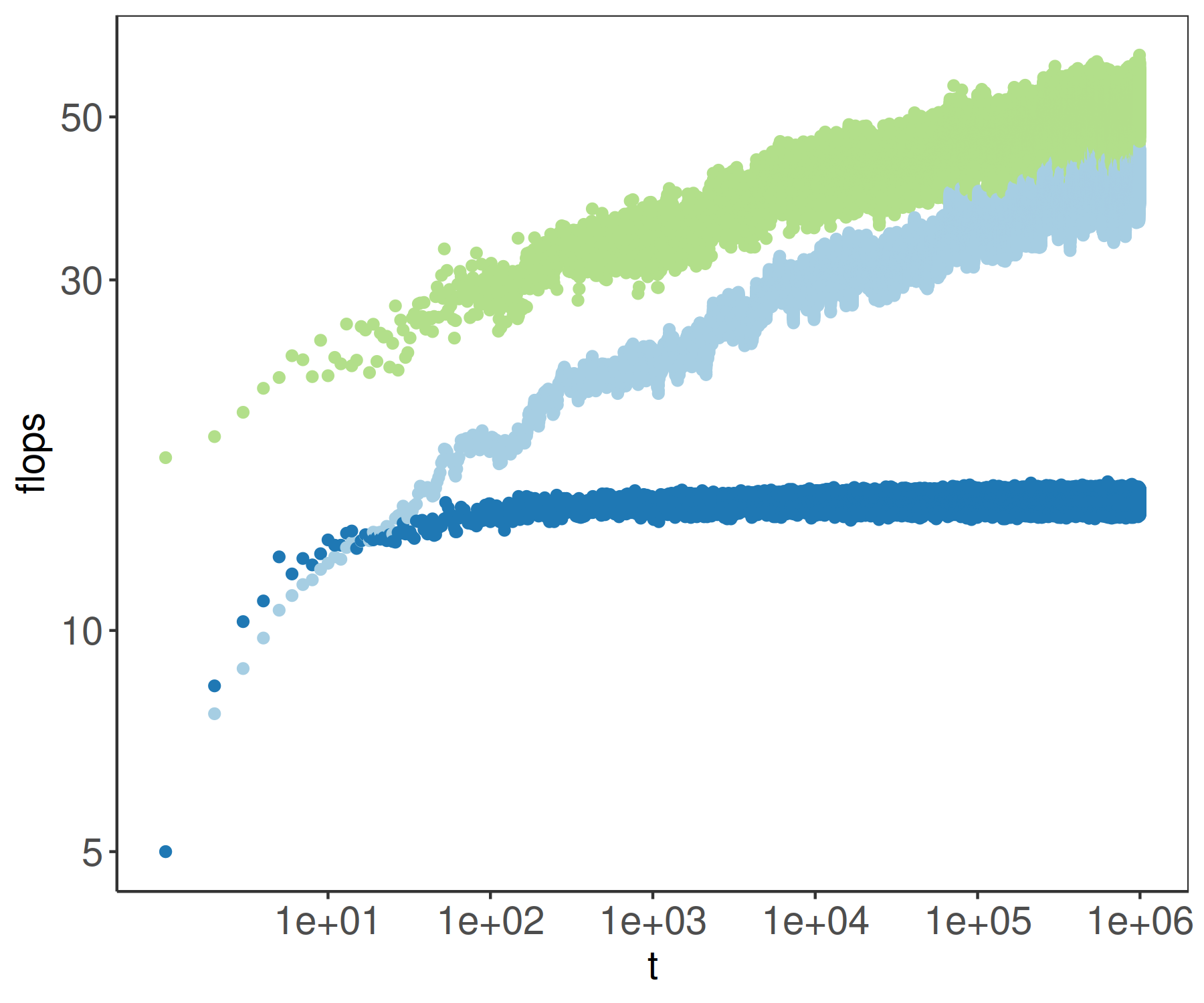}
		\caption{Flops per iteration in function of time for three FOCuS implementations. In green, the flops for FOCuS with pruning based on calculating the roots $l_1, \dots, l_n$ numerically. In light blue, FOCuS with the average value pruning. In blue, finally, FOCuS with the average value pruning and the adaptive maxima checking. Log-scale on both axes.}
		\label{fig:flops_per_time}
	\end{figure}

	In Figure \ref{fig:flops_per_time} we show the number of floating point operations as a function of time. The Figure was obtained by averaging results from 50 different sequences of length $1\times10^{6}$. Results were obtained under the Bernoulli likelihood. Under this likelihood, the cost for an update is negligible, given that this involves integer operations alone, and this allows for a better comparison of the costs of pruning and checking the maxima. We compare three different FOCuS implementations: (i) FOCuS with pruning based on the ordered roots $l_1, \dots, l_n$, where such roots are found numerically through the Newton-Raphson procedure, (ii) FOCuS with the average value pruning of Section \ref{sec:FOCuS_expo_fam} and lastly (iii) FOCuS with the average value pruning and the adaptive maxima checking of Section \ref{sec:apt_max_check}.
	
	We note that avoiding explicitly calculating the roots leads to a lower computational overhead when compared to Newton-Raphson. The best performances are, however, achieved with the addition of the adaptive maxima checking procedure, where we find a constant per iteration computational cost under the null centered around 15 flops per iteration. Without the adaptive maxima checking, the maximisation step is the most computationally demanding step of the FOCuS procedure, as  we need to evaluate $\mathcal{O}(\log(T))$ curves per iteration.
	
\begin{figure}[!htb]
		\centering
		\includegraphics[width=0.5\linewidth]{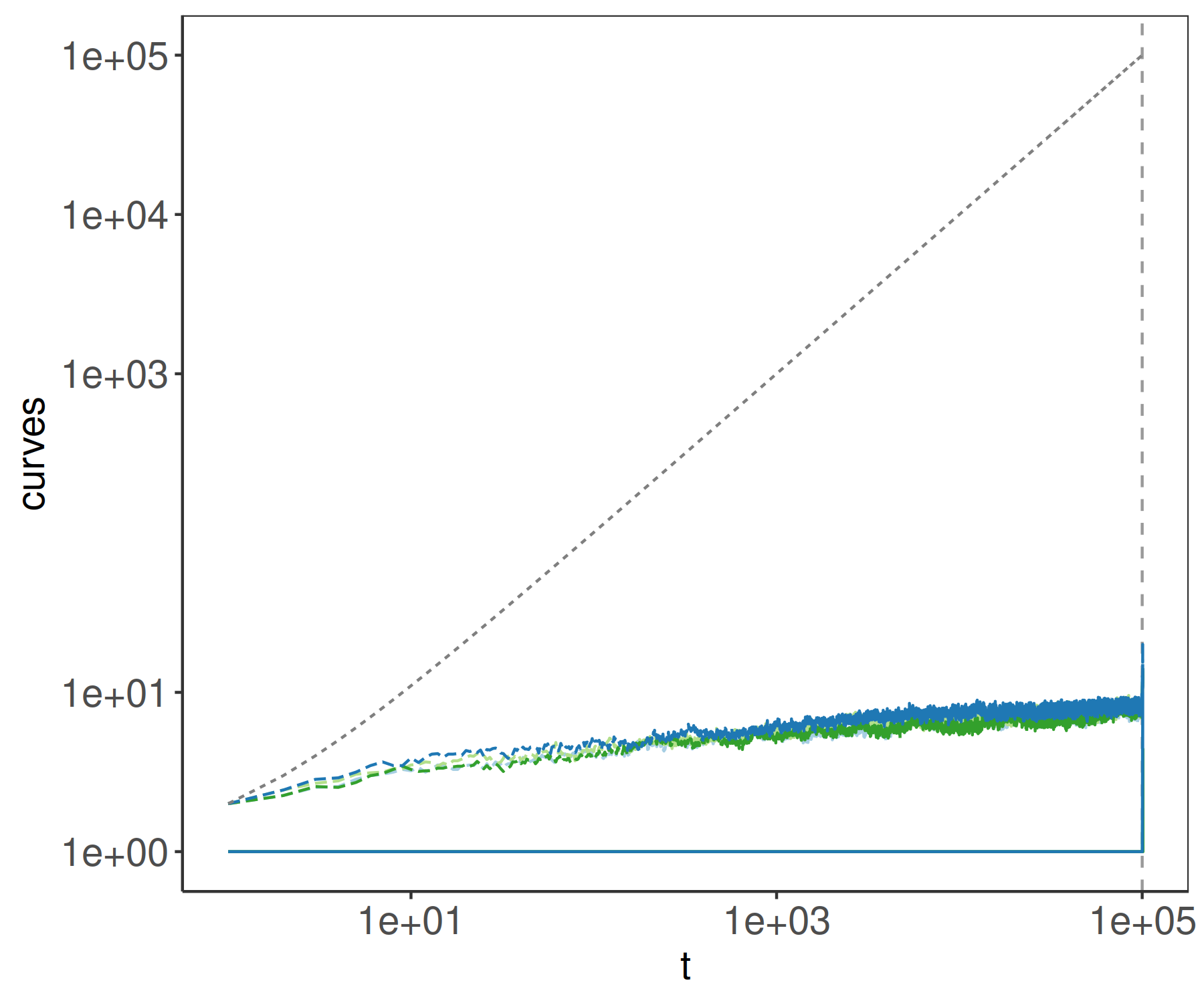}
		\caption{Number of curves to store and evaluations per iteration in function of time. The grey dotted line is the naive cost of computing the CUSUM likelihood ratio test. The dashed line are the number of curves stored by FOCuS over some Gaussian (light-green), Poisson (dark-green), Bernoulli (light-blue) and Gamma (dark-blue) realizations. The solid lines are the number of curves that need to be evaluated at each iteration with the adaptive maxima checking. Log-scale on both axes.}
		\label{fig:eval_per_time}
	\end{figure}

	In Figure \ref{fig:eval_per_time} we place a change at time $1\times10^{5}$ and we focus on the number of curves stored by FOCuS,  and the number of curves that need to be evaluated with the adaptive maxima checking. Furthermore, for comparison, we add a line for the naive cost of direct computation of the CUSUM likelihood-ratio test. We can see how, before we encounter a change, with the adaptive maxima checking routine we only need to maximise on average 1 curve per iteration, as compared to about 7.4 for the standard FOCuS implementation.
	After we encounter a change, then, the number of curves that need evaluation increases, as the likelihood ratio statistics increases and it is more likely to meet the condition of Proposition \ref{prop:2}. As it can be seen from the short spike after the change, this is only occurs for a short period of time preceding a detection. This empirically shows that FOCuS is $\mathcal{O}(1)$ computational complexity per iteration while being $\mathcal{O}(\log T)$ in memory, as we still need to store in memory on average $\mathcal{O}(\log T)$ curves.
	
	To illustrate the advantages of running FOCuS for the correct exponential family model, we consider detecting a change in variance in Gaussian data with known mean. We will assume that we have standardised the data so it has mean zero. A common approach to detecting a change in variance is to detect a change in mean in the square of the data \cite[]{inclan1994use}, so we will compare FOCuS for Gaussian change in mean applied to the square of the data against FOCuS for the Gaussian change in variance model (as in Table \ref{table:exponential_families}).
	
	
	For a process distributed under the null as a normal centered on 0 with variance $\theta_0 = 1$, we present 5 simulations scenarios for $\theta_1 = 0.75, 1.25, 1.5, 1.75$ and $2$. Each experiment consists of 100 replicates. Thresholds were tuned via a Monte Carlo approach to achieve an average run length of $1\times10^{5}$ under the null in the same fashion of \cite[Section 4.1]{chen2022high}. We then introduce a a change at time $1000$ and measure performances in terms of detection delay (the difference between the detection time and the real change).
	
\begin{figure*}[h]
		\centering
		\includegraphics[width=\linewidth]{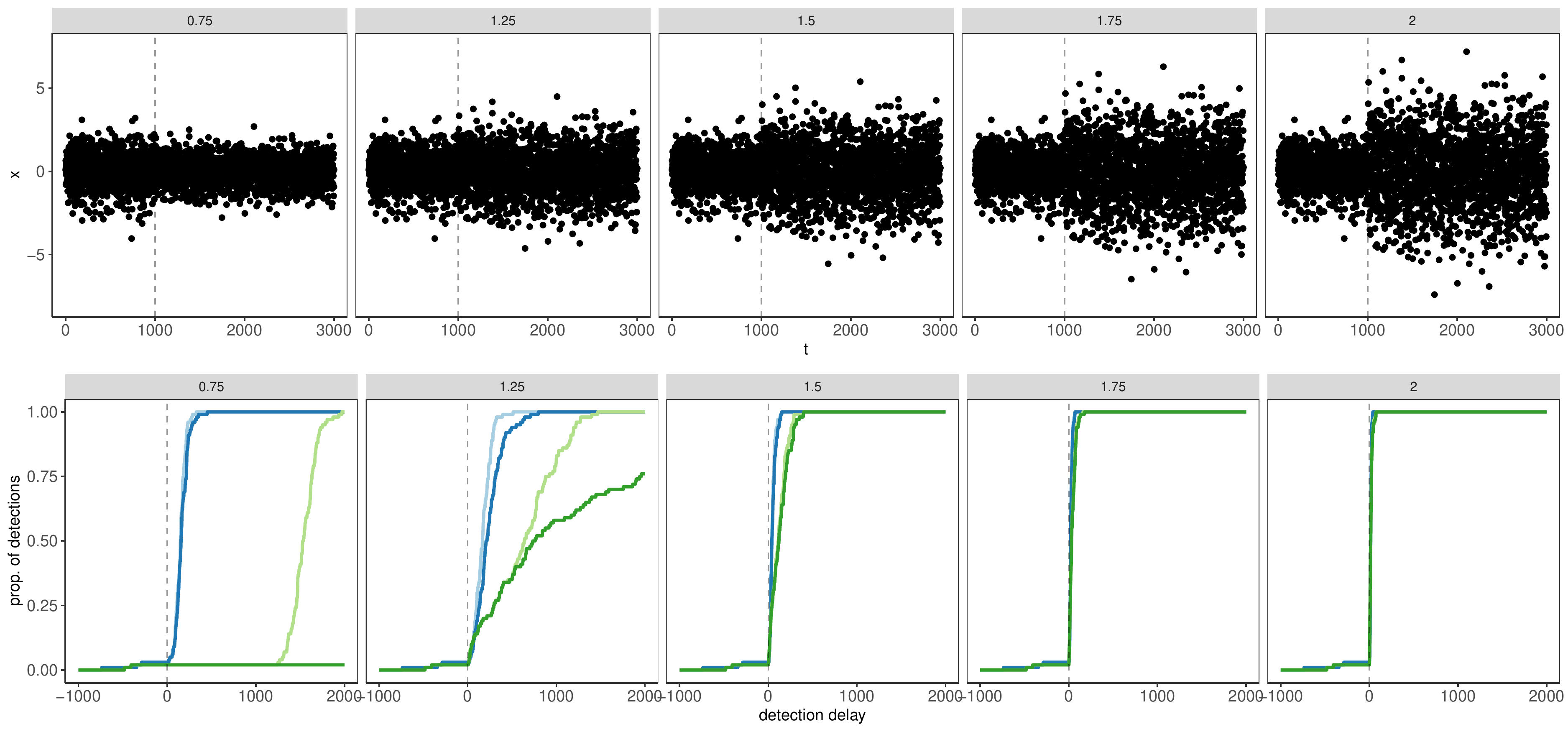}
		\caption{Empirical evaluation of FOCuS for Gaussian change-in-variance. Top row: example sequences for our simulation scenarios, with labels indicating the post-change parameter $\theta_1$, whilst vertical dotted line refers to the changepoint location $\tau$. Bottom row: proportion of detections as a function of the detection delay for Gaussian change-in-variance model with pre-change parameter known (light blue) and unknown (dark blue), and Gaussian change-in-mean applied to square of the data with pre-change parameter known (light green) and unknown (dark green). The vertical dotted line this time indicates the start of the change: the faster we get to 1 following the change, the better. Prior to the vertical line, we are essentially counting false positives.}
		\label{fig:variance_power}
	\end{figure*}

	In Figure \ref{fig:variance_power} we illustrate the scenarios and present results in terms of the proportion of detections within $t$ observations following the change.
	For a positive change large enough, e.g. for $\theta_1 = 2$, there is only a small advantage in employing the Gaussian change-in-variance model over the Gaussian change-in-mean applied to the square of the data. However, as we lower the signal-to-noise ratio and shift towards more subtle changes, we can see how using the correct model gives an increasing advantage in terms of reducing the detection delay. 

	\section{Discussion}
	
	We have presented an algorithm for online changepoint detection for one-parameter exponential family models that (i) exactly performs the likelihood-ratio test at each iteration; and (ii) empirically has a constant cost per-iteration. To the best of our knowledge, it is the first algorithm that achieves both of these. 
	
	The algorithm can only detect changes in a single parameter, and thus can only analyse univariate data. However this can provide the building block for analysing multivariate data. For example \cite{mei2010efficient} propose online monitoring multiple data streams by calculating statistics for a change for each individual data stream and then combining this information. There is an extensive literature on how one can combine such information in an efficient way \cite[for example][]{cho2015multiple,enikeeva2019high,tickle2021computationally}.
	
	A further challenge would be to extend the algorithm to deal with time-dependent data. Often methods that assume independence work well even in the presence of autocorrelation in the data providing one inflates the threshold for detecting a change \cite[]{lavielle2000least}. If the autocorrelation is strong, such a simple approach can lose some power, and either applying a filter to the data to remove the autocorrelation \cite[]{chakar2017robust} or adapting FOCuS to model it \cite[building on ideas in][]{romano2021detecting,cho2020multiple, hallgren2021changepoint} may be better.

	\section{Acknowledgments}
	
	This work was supported by the EPSRC grants EP/N031938/1 and EP/R004935/1, and BT as part of the Next Generation Converged Digital Infrastructure (NG-CDI) Prosperity Partnership.
	
	\bibliography{Bibliography}
	
	\newpage
	\appendix
	\section*{Appendices}
	\addcontentsline{toc}{section}{Appendices}
	
	\section{Proofs}
	\subsection{Deriving the Exponential family likelihood ratio}
	
	For an exponential family model of the form
	
	$${f(x\mid \theta )=\exp \!{\bigl [}\,\alpha (\theta )\cdot \gamma(x)-\beta(\theta )+\delta(x)\,{\bigr ]}},$$
	
	Our differences in likelihood are of the form
	
	\begin{eqnarray*}
		\lefteqn{\ell(x_{1:T} \vert\theta_0, \theta_1, \tau_i) - \ell(x_{1:T} \vert\theta_0, \theta_1, \tau_j)  = } \\
		& & [\alpha (\theta_1)-\alpha (\theta_0)]\sum_{t=\tau_i+1}^{\tau_j}\gamma(x_t)
		-[\beta(\theta_1)-\beta(\theta_0)](\tau_j-\tau_i).
	\end{eqnarray*}
	
	We have that
	\[
	\ell(x_{1:T} \vert\theta_0, \theta_1, \tau) := \sum_{t=1}^{\tau} \log f(x_t\vert\theta_0) + \sum_{t=\tau+1}^{T} \log f(x_t\vert\theta_1).
	\]
	
	Therefore, we have
	\begin{align*}
		& \ell(x_{1:T} \vert\theta_0, \theta_1, \tau_i) - \ell(x_{1:T} \vert\theta_0, \theta_1, \tau_j) =\\
		& = \sum_{t=\tau_i+1}^{\tau_j} \{\log f(x_t\vert\theta_1) - \log f(x_t\vert\theta_0)\}.
	\end{align*}
	
	
	Substituting in
	\begin{align*}
		& \log f(x_t\vert\theta_1) - \log f(x_t\vert\theta_0) = \\
		& = \left[\alpha (\theta_1)\cdot \gamma(x_t)-\beta(\theta_1)+\delta(x_t) \right] \\
		& \quad -\left[\alpha (\theta_0)\cdot \gamma(x_t)-\beta(\theta_0)+\delta(x_t) \right] \\
		& = [\alpha (\theta_1)-\alpha (\theta_0)]\gamma(x_t) - [\beta(\theta_1)-\beta(\theta_0)].
	\end{align*}
	
	
	gives the required result.

	\subsection{Ordering of roots determined by $\bar{\gamma}$ values}
	
	Define 
	\[
	\bar{\gamma}_{\tau_i:\tau_j}=\frac{1}{\tau_j-\tau_i} \sum_{t=\tau_i+1}^{\tau_j} \gamma(x_t)
	\]
	to be the average value of $\gamma(x_t)$ for $t=\tau_i+1,\ldots,\tau_j$, and define $\theta_1^\tau(\neq \theta_0)$ to be the root of
	\[
	\ell(x_{1:T} \vert\theta_0, \theta_1^\tau, \tau) - \ell(x_{1:T} \vert\theta_0, \cdot, T) = 0.
	\]
	
	\begin{proposition}
		Suppose that for our choice of $\theta_0$ the function
		
		\[\theta_1 : \rightarrow \frac{\beta(\theta_1)-\beta(\theta_0)}{\alpha(\theta_1)-\alpha(\theta_0)} \]
		
		is strictly increasing. Then the sign of $\bar{\gamma}_{\tau_i:\tau_j} - \bar{\gamma}_{\tau_j:T}$ is the same as the sign of $\theta_1^{\tau_i} - \theta_1^{\tau_j}$.
	\end{proposition}
	
	We have that
	
	\[
	[\alpha (\theta_1^{\tau})-\alpha (\theta_0)]\sum_{t=\tau+1}^{T}\gamma(x_t)
	-[\beta(\theta_1^{\tau})-\beta(\theta_0)](T-\tau) = 0.
	\]
	
	Rearrange this to form 
	
	\[
	\frac{\beta(\theta_1^{\tau})-\beta(\theta_0)}{\alpha (\theta_1^{\tau})-\alpha (\theta_0)} = \bar{\gamma}_{\tau:T}.
	\]
	
	By monotonicity, we have that $\theta_1^{\tau}$ is an increasing function of $\bar{\gamma}_{\tau:T}$. For $\tau_i < \tau_j<T$ we also have that
	
	\[
	\bar{\gamma}_{\tau_i:T} = \frac{T-\tau_j}{T-\tau_i}\bar{\gamma}_{\tau_j:T} + \frac{\tau_j-\tau_i}{T-\tau_i}\bar{\gamma}_{\tau_i:\tau_j},
	\]
	
	so the sign of $\bar{\gamma}_{\tau_i:\tau_j} - \bar{\gamma}_{\tau_j:T}$ is the same as the sign of $\bar{\gamma}_{\tau_i:T} - \bar{\gamma}_{\tau_j:T}$ because $\bar{\gamma}_{\tau_i:T}$ is a convex combination of $\bar{\gamma}_{\tau_i:\tau_j}$ and $\bar{\gamma}_{\tau_j:T}$. Putting this together gives the result.

	\subsection{Maxima checking bound}
	Define 
	\begin{eqnarray*}
		m_{\tau_i, \tau_j} = \max_{\substack{\theta_0\in H_0, \\ \theta_1}} \ell(x_{1:\tau_j} \vert\theta_0, \theta_1, \tau_i) - \max_{\theta_0\in H_0}\ell(x_{1:\tau_j} \vert\theta_0, \cdot, \tau_j),
	\end{eqnarray*}
	where $H_0$ denotes the set of possible values of $\theta_0$. $H_0$ will contain a single value in the pre-change parameter known case, or be $\mathbb{R}$ for the pre-change parameter unknown case.
	\begin{proposition}
		For any $\tau_1 < \tau_2 < ... < \tau_n < T$, we have 
		\[
		\max_{i=1,...,n} m_{\tau_i, T} \leq \sum_{i=1}^{n-1} m_{\tau_i, \tau_{i+1}} + m_{\tau_n, T}.
		\]
	\end{proposition}

	Denote by $\hat{\theta}_0^{\tau_i}$ the argmax of $\sum_{t=1}^{\tau_i} \log f(x_t\vert\theta_0)$ for $\theta \in H_0$. (Note that in the pre-change mean known case, we always have $\hat{\theta}_0^{\tau_i} = \theta_0$.)
	
	Now, consider the form of
	
	\begin{align*}
		m_{\tau_i, \tau_j} = & \sum_{t=1}^{\tau_i} \log f(x_t\vert\hat{\theta}_0^{\tau_i}) + \max_{\theta_1}\sum_{t=\tau_i+1}^{\tau_j} \log f(x_t\vert\theta_1)\\
		& - \sum_{t=1}^{\tau_j} \log f(x_t\vert\hat{\theta}_0^{\tau_j}).    
	\end{align*}
	
	Note the similarity of the first and third terms that will allow telescopic cancellations when summing the $m_{\tau_i, \tau_{i+1}}$. Setting $\tau_{n+1} := T$ for convenience, we have that for any $1 \leq k \leq n$,
	\begin{align*}
		& \sum_{i=1}^{n-1} m_{\tau_i, \tau_{i+1}} + m_{\tau_n, T} = \\
		& = \left[\sum_{t=1}^{\tau_1} \log f(x_t\vert\hat{\theta}_0^{\tau_1}) + \sum_{i=1}^{k-1}\max_{\theta_1}\sum_{t=\tau_i+1}^{\tau_{i+1}} \log f(x_t\vert\theta_1)\right] \\
		& \quad + \left[\sum_{i=k}^{n}\max_{\theta_1}\sum_{t=\tau_i+1}^{\tau_{i+1}} \log f(x_t\vert\theta_1)\right]  - \sum_{t=1}^{T} \log f(x_t\vert\hat{\theta}_0^{T}).
	\end{align*}
	
	
	We can compare this against
	\begin{align*}
		m_{\tau_k, T} = & \sum_{t=1}^{\tau_k} \log f(x_t\vert\hat{\theta}_0^{\tau_1}) + \max_{\theta_1}\sum_{t=\tau_k+1}^{T} \log f(x_t\vert\theta_1) \\
		& - \sum_{t=1}^{T} \log f(x_t\vert\hat{\theta}_0^{T}).
	\end{align*}
	noting that we have inequalities on the first two terms due to maximising the same likelihood over an expansion of the hypothesis set, and equality in the final term. This proves the result.
	
	The construction $\sum_{i=1}^{n-1} m_{\tau_i, \tau_{i+1}} + m_{\tau_n, T}$ is essentially fitting changepoints at every single one of the $\tau_i$. This compares against the construction $\max_{i=1,...,n} m_{\tau_i, T}$, which fits only one changepoint at the most promising $\tau_i$.
	
	Where $\{\tau_1, ...,  \tau_n\} \in \mathcal{I}_T$ and are therefore ordered in increasing/decreasing $\bar{\gamma}_{\tau_i:\tau_{i+1}}$ all representing up-changes/down-changes, it is the case that you don't gain much by fitting all of the $\tau_i$ as changepoints rather than just the best one. In the underlying data scenario of no change, the earlier $m_{\tau_i, \tau_{i+1}}$ will be very small, and it is $m_{\tau_n, T}$ that will contribute the most as it captures the fluctuations of recent events in the signal.

\end{document}